\documentclass[twocolumn,aps,floats,superscriptaddress,showpacs]{revtex4}
\usepackage{amsmath}

\usepackage{epsfig}
\usepackage{graphicx}
\usepackage{dcolumn}
\usepackage{bm}

\def\gapp{\lower.35em\hbox{$\stackrel{\textstyle>}{\sim}$}}
\def\lapp{\lower.35em\hbox{$\stackrel{\textstyle<}{\sim}$}}

\begin{document}
\bibliographystyle{apsrev}
%

\title{
Local defects and ferromagnetism in graphene layers.}
\author{M. A. H. Vozmediano}
\affiliation{
Unidad Asociada CSIC-UC3M,
Universidad Carlos III de Madrid, E-28911
Legan\'es, Madrid, Spain.}
\author{M. P. L\'opez-Sancho}
\affiliation{Instituto de Ciencia de Materiales de Madrid,
CSIC, Cantoblanco, E-28049 Madrid, Spain.}
\author{T. Stauber}
\affiliation{Instituto de Ciencia de Materiales de Madrid,
CSIC, Cantoblanco, E-28049 Madrid, Spain.}
\author{ F. Guinea}
\affiliation{Instituto de Ciencia de Materiales de Madrid,
CSIC, Cantoblanco, E-28049 Madrid, Spain.}
\affiliation{
Department of Physics, Boston University, 590
Commonwealth Avenue, Boston, MA 02215,USA }

\date{\today}
\begin{abstract}
We study the changes in the electronic structure induced by lattice defects
in graphene planes. In many cases, lattice distortions give rise to
localized states at the Fermi level. Electron-electron interactions lead to
the existence of local moments. The RKKY interaction between
these moments is always ferromagnetic, due to the semimetallic
properties of graphene.
\end{abstract}
%
%
%
%
\maketitle {\it Introduction.}
A number of recent experiments suggest that pure graphite
behaves as a highly correlated electron system\cite{Ketal02}.
In particular it shows a metal-insulator transition in magnetic fields and
insulating behavior in the direction perpendicular to the planes in different
samples\cite{Ketal00,Eetal02,KEK02,Ketal02,MHM02,Cetal02,Ketal03,Ketal03b}.
Recent results show ferromagnetic behavior\cite{Hetal03},
enhanced by proton bombardment\cite{Eetal03}, what opens up a new way to the
creation of organic magnets\cite{M02}.
In this paper we
study the formation of local moments near simple defects.
It is shown that many tyoes of lattice distortions, like cracks and
vacancies, can induce
localized states at the Fermi level, leading to the
existence of local moments. The RKKY interaction between
these moments is always ferromagnetic due to the
semimetallic properties of graphite. Hence, the RKKY
interaction does not lead to frustration and spin
glass features.

{\it The model.}
The conduction band of graphite is well described by a tight
binding model which includes the $\pi$ orbitals
which are perpendicular to the graphite planes at each
C atom\cite{SW58}. If the interplane hopping is
neglected, this model describes a semimetal, with zero
density of states at the Fermi energy, and where the
Fermi surface is reduced to two inequivalent K-points
located at the corner of the hexagonal
Brillouin Zone. The low-energy excitations with momenta
in the vicinity of the Fermi points have a linear dispersion
and can be described by a continuum model which reduces to the
Dirac equation in two dimensions\cite{SW58,M58,GGV93,GGV94,GGV96}.
The Hamiltonian density of the system is
\begin{align}
{\cal H}_{0}= \hbar v_F \int d^2 {\bf r} \bar{\Psi}({\bf r})
( i \sigma_x \partial_x + i \sigma_y \partial_y )
\Psi ({\bf r})\;,
\end{align}
where the
components of the two-dimensional spinor
$\Psi( {\bf \vec{r}} )=( \Psi_1 ( {\bf \vec{r}} ) , \Psi_2 ( {\bf \vec{r} }))$
correspond to the Bloch states of the two sublattices
in the honeycomb structure, independent in the absence
of interactions. In the clean two-dimensional system there is no
room for low energy electronic instabilities, the short range interactions
being irrelevant due to the vanishing density of states at the Fermi level.

It is known that disorder significantly changes the states
described by the two dimensional Dirac
equation\cite{CMW96,Cetal97,HD02}, and, usually, the density of
states at low energies is increased. Lattice defects, such as
pentagons and heptagons, or dislocations, can be included
in the continuum model by means
of a non-abelian gauge field\cite{GGV93,GGV01} that reproduces
the effects of the curvature of the lattice and the possible
exchange of Fermi points. Within the same theoretical scheme it
has also been shown that certain types of
disorder randomly distributed in the graphene lattice
enhances the effect of the interactions\cite{SGV05} and can
stabilize new phases. In addition, a
graphene plane can show states localized at
interfaces\cite{WS00,W01}, which, in the absence of other types of
disorder lie at the Fermi energy. Structures
with mixed $sp^2$ and $sp^3$ bonding
can also lead to localized states\cite{OS91}.

The tight binding model defined by the
$\pi$ orbitals at the lattice sites can have edge states when the
sites at the edge belong all to the same
sublattice\cite{WS00,W01,Hetal04}(zigzag edge). These states lie at zero
energy which for neutral graphene planes correspond to the
Fermi energy.
In the continuum model described
earlier localized states are normalizable solutions
 $( \Psi_1 ( {\bf \vec{r}} ) , \Psi_2 ( {\bf \vec{r} }))$
 of the Dirac equation:
\begin{eqnarray}
\left( \partial_x + i \partial_y \right) \Psi_1 ( {\bf \vec{r}} )
&= &0 \nonumber
\\ \left( \partial_x - i \partial_y \right) \Psi_2 ( {\bf \vec{r}} ) &= &0
\end{eqnarray}
These equations are satisfied if $\Psi_1 ( {\bf \vec{r}} )$ is an
analytic function of $z = x + i y$ and $\Psi_2 ( {\bf \vec{r}} ) =
0$, or if $\Psi_1 ( {\bf \vec{r}} ) = 0$ and $\Psi_2 ( {\bf
\vec{r}} )$ is an analytic function of $\bar{z} = x - i y$.
Zigzag edge states can be obtained as follows.
Consider a semi-infinite honeycomb lattice with an edge at $y=0$
and which occupies the half plane $x>0$. A possible solution which
decays as $x \rightarrow \infty$ is
$$\Psi_1 ( x , y ) \propto e^{-
k z} = e^{i k y} e^{- k x} , \Psi_2 ( {\bf \vec{r}} ) = 0\;.$$
 These
solutions satisfy the boundary conditions at $y=0$ if the last
column of carbon atoms belong to the sublattice where the
component $\Psi_1$ is defined. Then, the next column belongs to
the other sublattice, where the amplitude of the state is, by
construction, zero.
\begin{figure}
  \begin{center}
    \epsfig{file=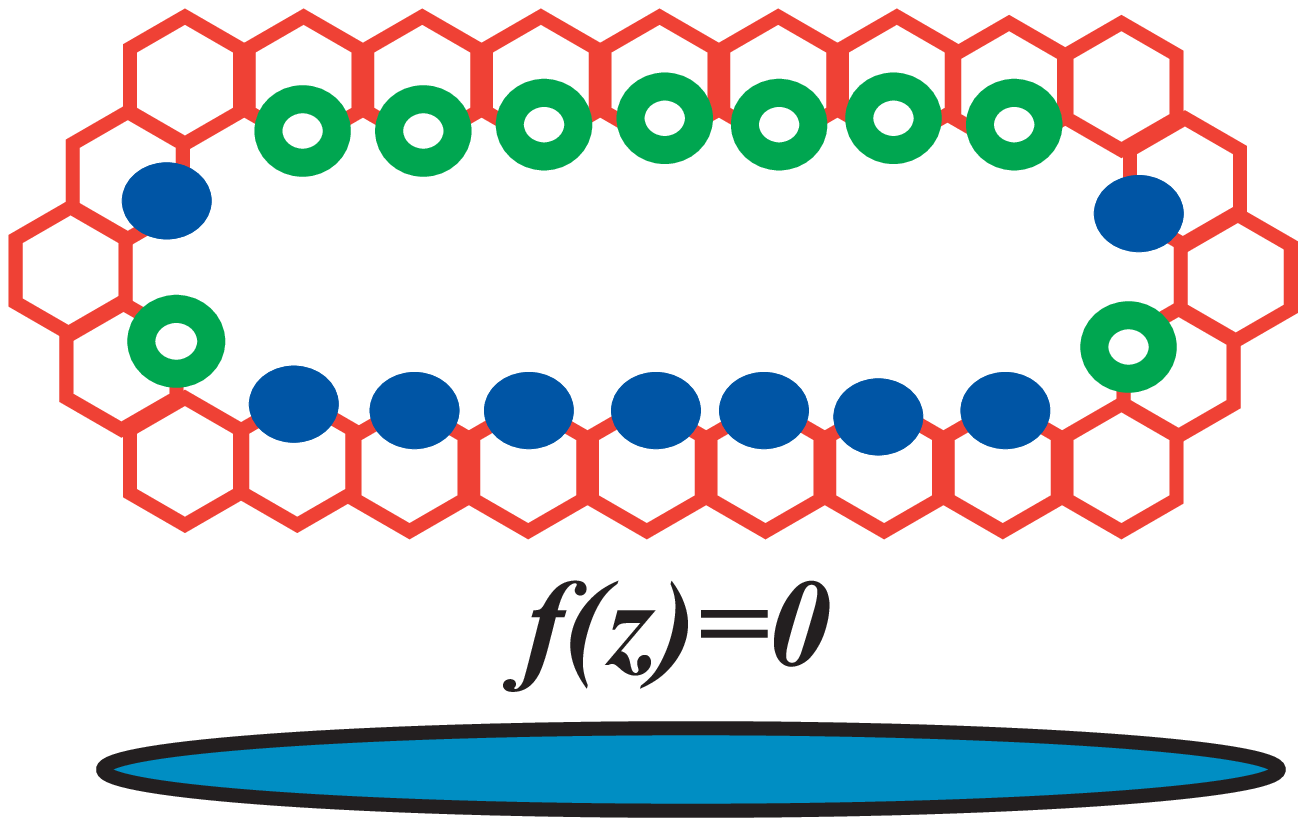,height=5cm}
    \caption{Elongated crack in the honeycomb structure. The crack is such
    that the sites in the upper edge belong to one sublattice, while those
    at the lower edge belong to the other. Bottom: approximate cut in the
    complex plane which can be used to represent this crack at long distances.}
    \label{crack}
\end{center}
\end{figure}

This kind of  solutions can be generalized to describe other types of extended defects
that will be produced in experiments where graphite samples are bombarded by protons.
In a strongly disordered sample, large defects made up of many
vacancies can exist. These defects will give rise to localized states,
when the termination at the edges is locally similar to the
surfaces discussed above. Note that, if the bonds at the edges are
saturated by bonding to other elements, like hydrogen, the states
at these sites are removed from the Fermi energy, but a similar
boundary problem arises for the remaining $\pi$ orbitals.
The only possible localized
states can exist at zero energy, where the density of extended
states vanishes. The wave functions obtained from the Dirac
equations will be normalizable and analytic functions of the variables
$z =x+iy$ or $\bar{z} = x-iy$ of the form
$$\Psi ( z ) \equiv [ f ( z ) , 0 ]$$
obeying the boundary conditions
imposed by the shape of the defect.

Extended vacancies with approximate
circular shape can support solutions of the type
$$f( z ) \propto z^{-n} , n > 1\;.$$
By using conformal mapping
techniques, solutions can be found with the  boundary conditions
appropriate to the shapes of different defects.

A simple case is the elongated crack depicted in
Fig.[\ref{crack}], which we assume to extend from $x=-a$ to $x=a$, and to have
a width comparable to the lattice constant along the $y$ axis. The
analytic function $f ( z )$ associated with localized states
near a crack of this shape should satisfy ${\rm Re} f ( z ) = 0$
at the crack edges, because the boundaries of the crack include
atoms from the two sublattices. Hence, the boundary of the crack
leads to a branch cut in the complex function $f ( z )$. Labelling
edge states by a quantum number $n$, we find that the function
$\Psi$ can be written for these states as:
$$\Psi_n \equiv \left\{
{\rm Re} \left[ \frac{A}{z^n \sqrt{z^2 - a^2}} \right] , 0
\right\}\;.$$
 A similar solution is obtained by exchanging the upper
by the lower spinor component, and replacing $z \leftrightarrow
\bar{z}$. Because of the discreteness of the lattice, the allowed
values of $n$ should be smaller than the number of lattice units
spanned by the crack.

\begin{figure}
\begin{center}
\epsfig{file=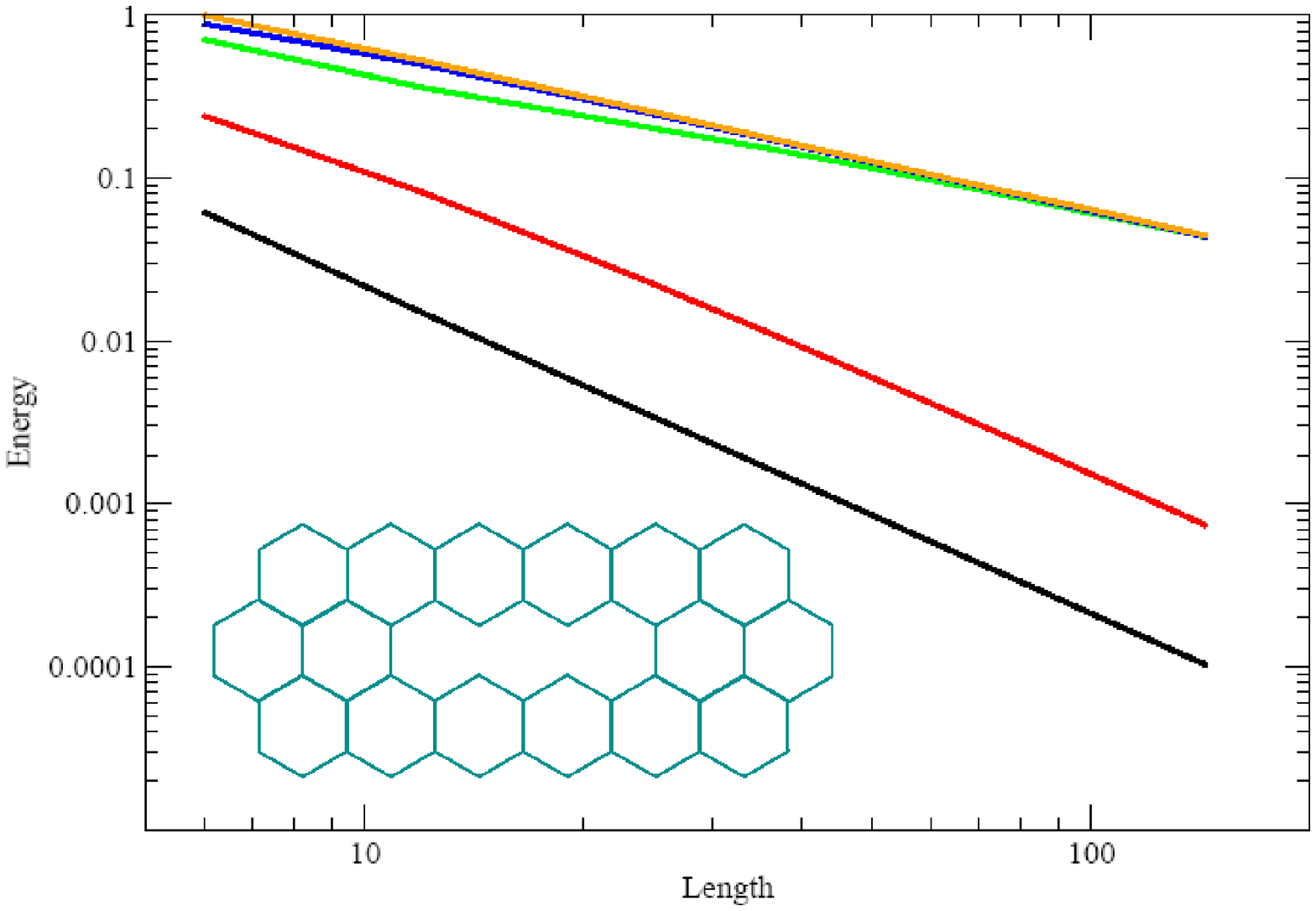,height=7cm}
\caption{Energy of the five states with positive energy
closest to $\epsilon = 0$ for a cluster of size $L \times L$
which contains at the center the defect
shown in the inset. Periodic boundary conditions are used. The spectrum is
symmetric around $\epsilon = 0$.}
\label{scaling}
\end{center}
\end{figure}

We have checked numerically the existence of these
localized  states by diagonalizing
the tight binding hamiltonian in finite lattices of different sizes. The
dependence of some of the states close to the chemical potential (zero
energy) on the cluster size is shown in Fig.[\ref{scaling}]. The
delocalized states show a dependence $\epsilon_{\rm del} \propto L^{-1}$, consistent
with the properties of the Dirac equation from which they can be derived.
The states closest to $\epsilon = 0$, show a dependence
 $\epsilon_{\rm loc} \propto L^{-2}$, which suggest a power law localization,
 in agreement with the previous analysis.

These states are half filled in a neutral graphene plane.
In the absence of electron interactions, this leads
to a large degeneracy in the ground state. A finite local repulsion
will tend to induce a ferromagnetic alignment of the electrons
occupying these states, as in similar cases with degenerate
bands\cite{Vetal99}.

\begin{figure}
  \begin{center}
   \epsfig{file=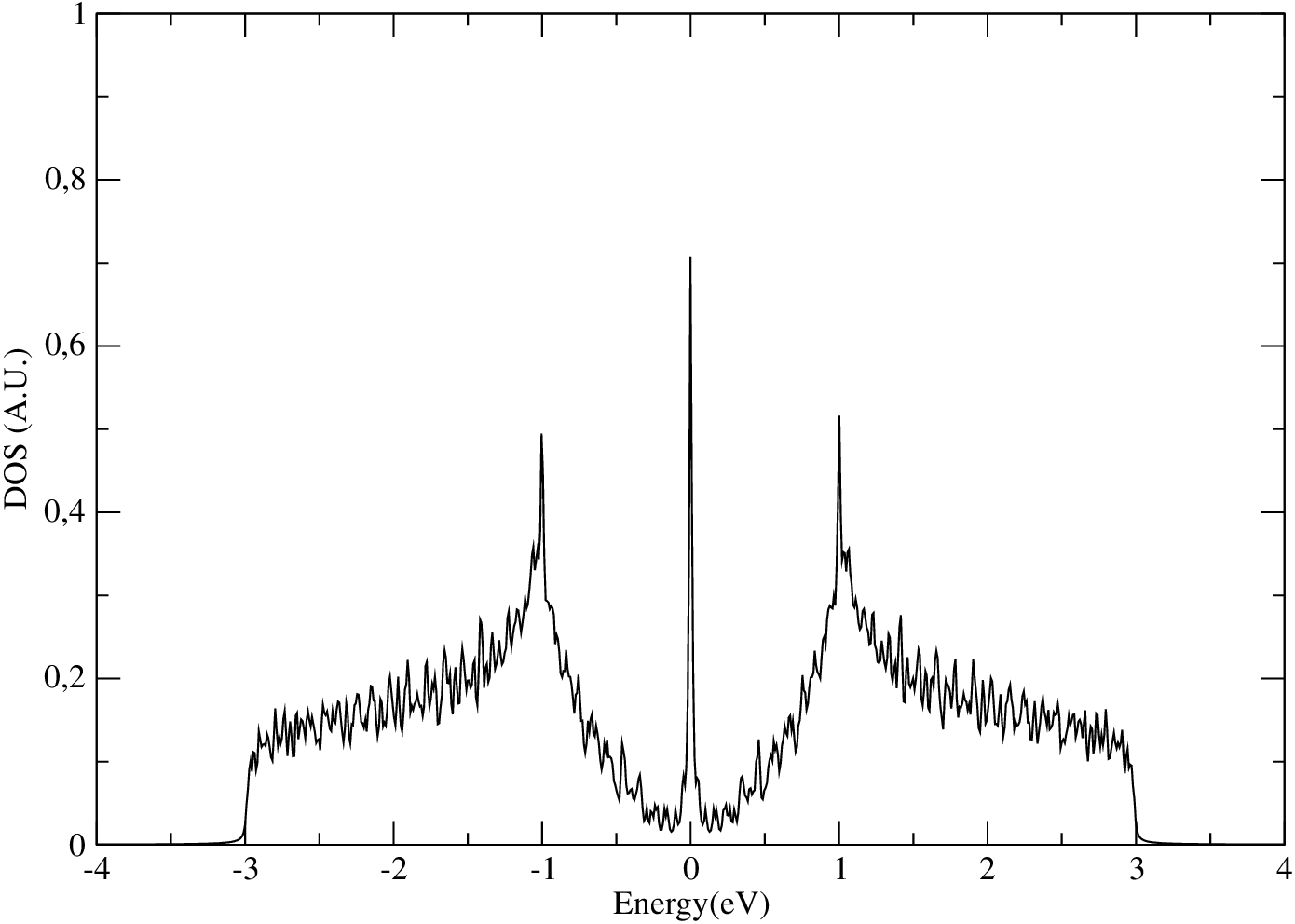,height=5cm}
    \caption{Density of states of a $26 \times 26$ cluster with
a large defect with zig-zag boundary as the one shown in Fig.[\ref{crack}]
The on-site interaction term is
$U = 0$.}
    \label{dos1}
\end{center}
\end{figure}

\begin{figure}
  \begin{center}
   \epsfig{file=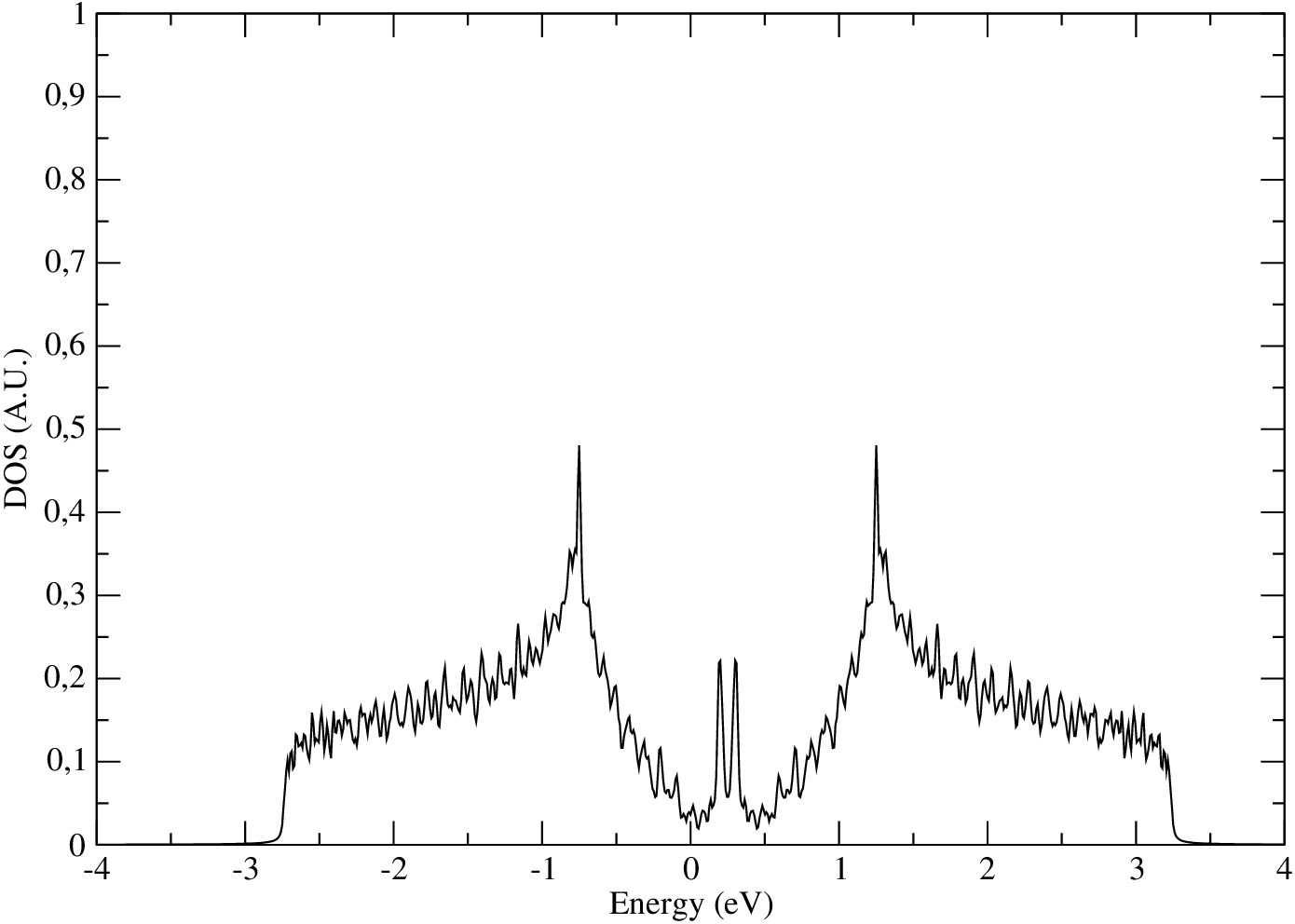,height=5cm}
    \caption{Same as Fig.[\ref{dos1}] with an on-site interaction
    $U=0.5t$.}
    \label{dos2}
\end{center}
\end{figure}
To verify this behavior and to compare with previous results
we have performed an unrestricted
Hartree Fock calculation of the graphene cluster with
different types of defects including
large vacancies as the one shown in Fig.[\ref{crack}].
Similar calculations done in  graphene
ribbons   have shown
non-bonding molecular orbitals localized mainly along the
zigzag edges (edge states)\cite{Yetal04,YSHW03}. These edges
 have partly flat bands which give rise to
 a sharp peak in the density of states at the Fermi energy.
A finite on-site Coulomb interaction
leads to an instability  of these flat bands
and to the existence of spin-polarized  states.
We have verified that the similar behavior
can also be obtained near the boundary of a large defect as the
one shown in Fig.[\ref{crack}]. We have performed calculations in
a honeycomb lattice of various sizes with a maximum of 26$\times$26 unit cells,
i. e. a cluster of 52$\times$ 52 sites.
In the absence of defects the density of states is zero at the Fermi level.
Figure [\ref{dos1}] shows the charge density distribution of
a  lattice with an extended vacancy   shaped as shown in Fig.[\ref{crack}],
without on-site interaction. The Fermi level is located at zero. A peak in the
Fermi level is formed due to the presence of the defect. The total
magnetization of the cluster is zero. When the on-site interaction is
switched on, a spin polarization arises of a magnitude proportional
to the size of the defect.  Fig.[\ref{dos2}] shows  the density profile of
the polarized cluster obtained with $U=0.5t$.

Next we will  analyze the influence of
 these magnetic moments on the conduction band.
The hopping between the states involved in the formation of these moments
and the delocalized states in the conduction band vanishes by definition,
if the localized states lie at zero energy. Hence, a Kondo-like coupling
mediated by the hopping will not be induced. The localized electrons act as a
 reservoir of localized moments, which interact
with the valence electrons via the on site Hubbard repulsion, $U$.
The wavefunction of the localized electrons overlaps on many sites with the
valence electrons. On these sites, the valence electrons will tend to be
polarized with their spins parallel to the spins of the localized electrons.
The change in the energy at a given site for the valence electrons near
the defect is of order $U$, distributed over $N$ sites.
Then, the  valence electrons will
induce long range RKKY interactions between the localized moments, which can
 be estimated by adding the contributions from all sites. We find:

\begin{equation}
J_{RKKY} ( {\bf \vec{r}} ) \sim
U^2 N^2  \int
d^2 {\bf k} e^{i {\bf \vec{k}} {\bf \vec{r}}
} \chi ( {\bf \vec{k}} ) \sim U^2 N^2 \frac{a^4}{v_F | {\bf \vec{r}} |^3}
\label{RKKY}
\end{equation}
 Where $a$ is the lattice constant, and the static susceptibility is given by:
 $\chi ( {\bf \vec{k}} ) \propto | {\bf \vec{k}} |$\cite{GGV96},
and $a$ is the lattice constant.

Due to the absence of a finite Fermi surface, the RKKY interaction
in eq.(\ref{RKKY}) does not have oscillations. Hence, there are no
competing ferro- and antiferromagnetic couplings, and the
magnetic moments will tend to be ferromagnetically aligned. The total
polarization per unit area at low temperatures is proportional to
$N c$, where $c$ is the concentration of defects, and $N$ is proportional to
the average size.

We can make an estimate of the Curie temperature from the coupling between
the local magnetic moments given in eq.(\ref{RKKY}). The entropy cost of
aligning ferromagnetically moments is $S \sim T$ per
moment. The average distance between moments is $| {\bf \vec{r}}  | \sim
c^{-1/2}$. Hence, the free energy per moment in the ferromagnetic phase
can be written as:
\begin{equation}
{\cal F} ( m ) = \left( - c_1 \frac{U^2 N^2 a^4 c^{3/2}}{v_F} + c_2
T \right) m^2 + \cdots
\label{free_en}
\end{equation}
where $c_1$ and $c_2$ are numerical constants of order unity.
The value of the free energy will be negative (and below the value in the
paramagnetic phase) at a Curie temperature given by:
\begin{equation}
T_c \sim \frac{U^2 N^2 a^4 c^{3/2}}{v_F} \approx \frac{U^2}{W} \frac{N
  a^3}{l^3}
\label{TC}
\end{equation}
where $W$ is the conduction electron bandwidth,$W \sim v_F / a$,
and $l$ is the average distance between impurities.

The Curie temperature depends on the concentration and size of the
defects. Assuming, as an example, that $N \sim 10$ and $l \sim 10^2 a$, we
obtain a saturation magnetization of $10^{-3}$ Bohr magnetons per unit cell,
and a Curie temparature $T_C \sim 10^{-4} U^2 / W$. The value of $U^2 / W$
can be estimated to be $\sim 1$eV. Then, these arguments give $T_C \sim
1$K. This temperature is considerably lower than the experimentally
observed ones. It is worth noting, however, that this analysis does not take
into account the enhancement of the susceptibility of the conduction
electrons, percolation effects due to the random distribution of impurities,
and the finite extension of the localized states induced by the defects.

{\it Conclusions.}
We have shown that, under very general circumstances, lattice defects,
vacancies and voids in the graphene structure
give rise to localized states at the Fermi energy. The number of these states
scales roughly with the perimeter of the defect. Repulsive electron-electron
interactions lead to the polarization of these states, and to the formation
of local moments. The RKKY interaction mediated by the valence electrons
decays as $r^{-3}$, where $r$ is the distance between defects, and shows no
oscillations, due to the vanishing of the Fermi surface in a graphene
layer. The interaction is ferromagnetic, and the system cannot show the
frustration effects and spin glass features observed in other disordered
systems with local moments. On the other hand, the Curie temperature
estimated assuming a random distribution of local moments is low, $T_{\rm C} \sim
1$K, for reasonable values of the defect concentration. It may happen that
percolation effects, and the finite extension of the localized states which
give rise to the local moments will increase the value of $T_{\rm C}$.

{\it Acknowledgments.}
We thank P. Esquinazi for sharing his experiments
with us and for many illuminating discussions.
Funding from MCyT (Spain) through grant MAT2002-0495-C02-01 is
acknowledged. One of us (F. G:) would like to thank the hospitality of Boston
University, where part of this work was done.
\bibliography{graphite_21}
\end{document}